\documentclass{report}
\usepackage[english]{babel}
\usepackage{tikz}

\usepackage[letterpaper,top=2cm,bottom=2cm,left=3cm,right=3cm,marginparwidth=1.75cm]{geometry}

\usepackage{xcolor}  
\usepackage{graphicx}
\usepackage{booktabs}
\usepackage{colortbl}
\usepackage{float} 
\graphicspath{{images/}} 

\definecolor{lightgray}{gray}{0.9}

\usepackage{amsmath}
\usepackage{graphicx}
\usepackage{placeins}
\usepackage{float}
\usepackage{subfigure}

\usepackage[colorlinks=true, allcolors=blue]{hyperref}
\usepackage{listings}

\definecolor{codegreen}{rgb}{0,0.6,0}
\definecolor{codegray}{rgb}{0.,0.5,0.5}
\definecolor{codepurple}{rgb}{0.047, 0.318, 0.51}
\definecolor{backcolour}{rgb}{1, 1, 1}

\lstdefinestyle{mystyle}{
    backgroundcolor=\color{backcolour},   
    commentstyle=\color{codegreen}, 
    keywordstyle=\color{magenta},
    numberstyle=\tiny\color{codegray},
    stringstyle=\color{codepurple},
    basicstyle=\ttfamily\footnotesize,
    breakatwhitespace=false,          
    breaklines=true,                 
    captionpos=b,                    
    keepspaces=true,                 
    numbers=left,                    
    numbersep=5pt,                  
    showspaces=false,                
    showstringspaces=false,
    showtabs=false,                  
    tabsize=2 
}
\usepackage{ifxetex}
\usepackage{ifluatex}
\newif\ifxetexorluatex 
\ifnum 0\ifxetex 1\fi\ifluatex 1\fi>0
   \xetexorluatextrue
\fi

\ifxetexorluatex
  \usepackage{fontspec}
\else
  \usepackage[T1]{fontenc}
  \usepackage[utf8]{inputenc}
  \usepackage[lighttt]{lmodern}
\fi

\usepackage{textcomp}
\usepackage{xcolor}
\usepackage{listings}
\usepackage{upquote}

\definecolor{keyword}{HTML}{2771a3}
\definecolor{pattern}{HTML}{b53c2f}
\definecolor{string}{HTML}{be681c}
\definecolor{relation}{HTML}{7e4894}
\definecolor{variable}{HTML}{107762}
\definecolor{comment}{HTML}{8d9094}

\lstdefinelanguage{cypher}
{
	morekeywords={
		CALL, YIELD, MATCH, OPTIONAL, WHERE, NOT, AND, OR, XOR, RETURN, DISTINCT, ORDER, BY, ASC, ASCENDING, DESC, DESCENDING, UNWIND, AS, UNION, WITH, ALL, CREATE, DELETE, DETACH, REMOVE, SET, MERGE, SET, SKIP, LIMIT, IN, CASE, WHEN, THEN, ELSE, END,
		INDEX, DROP, UNIQUE, CONSTRAINT, EXPLAIN, PROFILE, START,
	}
}

\newcommand{\mycdots}{\cdot\!\cdot\!\cdot}
\title{\textbf{Predictive Query-based Pipeline for Graph Data}}
\author{Plácido Antonio Souza Neto \\
\\ \textit{LIFO - Laboratoire d'Informatique Fondamentale d'Orléans}
\\ \textit{Université d'Orléans}
\\ \\ \textbf{Postdoc REPORT}}

\begin{document}
\maketitle  
    

\chapter{Introduction}

In the context of the DOING project, funded by APR-IA and based in the Val de Loire region of France, 
we've been developing a methodology (pipeline) to extract meaningful information from graph data. 
So, this work was funded by the DOING project.
Our primary goal is to structure this information into a graph database, creating a knowledge graph 
that can be intelligently manipulated. To ensure practical application, we've focused on the health sector 
as our domain of interest.

The DOING project is dedicated to developing data science queries that can facilitate informed 
decision-making by specialist professionals. By combining expertise from Automatic Language Processing, Databases, 
and Artificial Intelligence, we aim to transform raw data into valuable information and ultimately, actionable knowledge.

The DOING project, originating from the RTR-DIAMS and GDR-MADICS working groups,
aims to develop innovative methods, algorithms, and tools for transforming data into 
actionable information and knowledge. Within Task 2, ``\textit{Data Science Queries}'', 
we are specifically focused on creating an approach for manipulating and maintaining databases 
using machine learning and query techniques.

The contribution of this work is to propose a upcoming method/pipeline which helps to answer 
queries from specialists. The main goal is to, given a graph database schema and its instance, 
propose a set of queries to help specialists to answer their questions.

The remainder of this document is structured as follows. Chapter \ref{chap:embedding} explores various 
approaches for graph embedding generation, including GraphSAGE, Node2Vec, and FastRP. Graph embedding is a 
valuable technique for representing complex graph data in a lower-dimensional space. By transforming graphs into vectors, 
it facilitates the analysis and processing of large-scale datasets. Chapter \ref{chap:similarity_embedding} presents 
a straightforward method for analyzing the similarity between nodes using graph embeddings. 
The similarity results are visualized through t-SNE, Isomap, Spectral Embedding, and MDS techniques. 
Chapter \ref{chap:predictive_query} introduces a predictive query-based pipeline for graph data. 
The pipeline comprises some steps which includes graph embedding generation, 
similarity analysis, query definition, visualization and results.

 \chapter{Embedding for Graphs Representation}  
\label{chap:embedding}

Graphs face challenges when dealing with massive datasets. 
They are essential tools for modeling interconnected data and often become computationally expensive~\cite{skienna08,Sedgewick}. 
Graph embedding techniques~\cite{Xu2021}, on the other hand, provide an efficient approach. 
By projecting complex graphs into a lower-dimensional space, these techniques simplify the analysis and 
processing of large-scale graphs.
  
Graph embedding is a method that maps complex networks or graphs into a lower-dimensional vector space~\cite{khoshraftar2022surveygraphrepresentationlearning}. 
This transformation simplifies the analysis of datasets while preserving their structure and meaning. 
From representing nodes as points in a continuous space [vectors], it is possible to run machine learning (ML) techniques. 
The ML techniques include clustering, classification, recommendation or link prediction.

There are different approaches to graph representation through embedding. Each approach uses a different technique, because there are some key challenges, like heterogeneity or dimensionality. 
For instance, there are approaches that consider a random walk-based methods or graph neural network-based methods. The most used approaches for graph embedding are: GraphSAGE~\cite{HamiltonYL17}, FastRP~\cite{fastrp} and Node2Vec~\cite{grover2016node2vec}.   

GraphSAGE~\cite{HamiltonYL17} (Graph SAmple and Aggregate) is a scalable framework for learning node embeddings in large-scale graphs. 
It's designed to handle data with millions of nodes, making it a important tool for graph applications. 
GraphSAGE learns node embeddings by sampling and aggregating information from a node's neighbors. 
This approach allows it to generalize to unseen nodes, making it more scalable and adaptable than traditional methods.
Its ability to sample and aggregate neighbor information makes it scalable and generalizable. 

Node2Vec~\cite{grover2016node2vec} is a graph embedding technique that learns low-dimensional representations of nodes in a graph. 
These representations can be used for different tasks, such as node classification, link prediction, and community detection.
By considering both local and global structures, it can capture information for different graph-related tasks.

FastRP~\cite{fastrp} (Fast Random Projection) is a graph embedding technique that offers a significant speedup compared to other methods like DeepWalk and node2vec. It's particularly effective for large-scale graphs.
It also leverages random projections to efficiently reduce the dimensionality of a graph's similarity matrix. This matrix captures the pairwise similarities between nodes, and by projecting it onto a lower-dimensional spaces.
This approach is a scalable and efficient graph embedding method that offers a compelling alternative to traditional approaches. 
Its simplicity and speed make it a interesting tool for analyzing large-scale graphs.

Considering the embedding approaches, there are some graph database management system that generate embedding representation for graphs. It is the case of Neo4J~\cite{vukotic2015neo4j}. 
Neo4J is a graph database management system (DBMS) that is specifically designed to store and manage data in the form of graphs. 
Unlike traditional relational databases, which store data in tables, Neo4j stores data as nodes (entities) and relationships (connections between entities). 
This makes it particularly well-suited for handling complex, interconnected data.

\begin{figure} 
    \centering
    \begin{tikzpicture}
        \node[circle,draw] (Bob) at (0,0) {Bob}; 
        \node[circle,draw] (Alice) at (3,0) {Alice};
        \node[circle,draw] (James) at (6,0) {James};
        \node[circle,draw] (MovieA) at (1.5,2) {The Hobbit};
        \node[circle,draw] (MovieB) at (1.5,-2) {The LFR};
        \node[circle,draw] (MovieC) at (4.5,2) {Anatomy};
        \node[circle,draw] (MovieD) at (4.5,-2) {MotherInstinct};
      
        \draw[->,ultra thick] (Bob) -- (MovieA) node[midway,above] {5};
        \draw[->, thick] (Bob) -- (MovieB) node[midway,below] {3};
        \draw[->, very thick] (Alice) -- (MovieB) node[midway,right] {4};
        \draw[->, dashed] (Alice) -- (MovieC) node[midway,above] {2};
        \draw[->, dotted] (James) -- (MovieD) node[midway,below] {1};
        \draw[->, very thick] (James) -- (MovieC) node[midway,right] {4};
      
        \node[draw, rounded corners=2pt, align=left] at (10,0) {
          \textbf{Edge Styles:}\\
          * Ultra thick: 5 stars\\
          * Thick: 4 stars\\ 
          * Very thick: 3 stars\\
          * Dashed: 2 stars\\
          * Dotted: 1 star
        };
      \end{tikzpicture}
      \caption{Example of a \textit{MovieLens} Graph. \label{fig:movie_graph}}
    \end{figure}
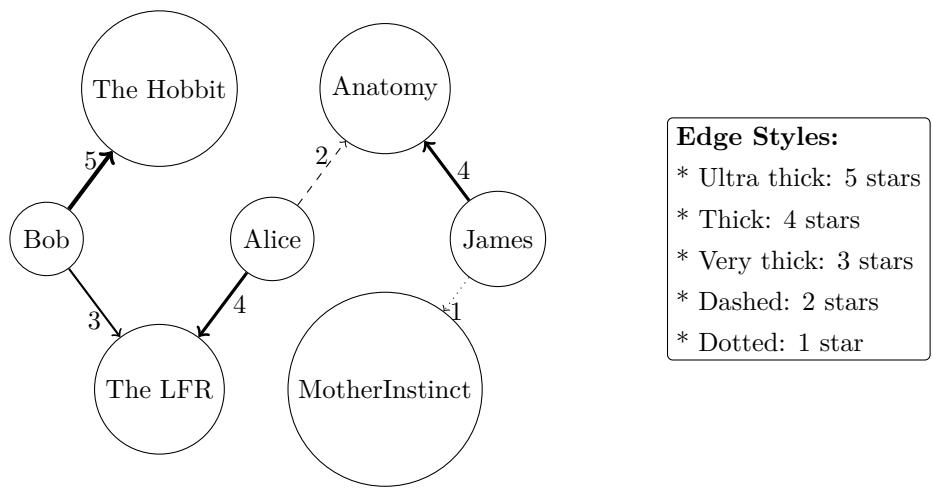

Consider the \textit{MovieLens} graph example as presented in Figure~\ref{fig:movie_graph}. 
\textit{Nodes} represent movies and users, and \textit{Edges} connect users to the movies they've rated.
The weight of an \textit{Edges} might represent the rating (e.g., 1 to 5 stars). 

\begin{itemize}
    \item Node: Movie (e.g., ``The Hobbit'', ``The Lord of the Rings'' , ``Anatomy of a Fall'', ``Mother's Instinct'');  
    \item Node: User (e.g., Bob, Alice, James, Anne).
    \item Edge: Bob rated ``The Hobbit" with 5 stars.
    \item Edge: Alice rated ``The Lord of the Rings'' with 4 stars.
\end{itemize}

A possible embedding space  might look like a N-dimensional space, generated by one of the cited embedding approaches.
Using an embedding algorithm like Node2Vec, it will performs random walks on the graph to learn representations of nodes Movie and User, 
and the type of relationship between them. Nodes that are frequently visited together during random walks will have similar embeddings.
While it is not easy to visualize a N-dimensional space directly, it is important to apply a 2D reduction algorithm for a simplified representation.

A user (e.g. Anne or Bob) who frequently rates movies like ``Anatomy of a Fall'' and ``Mother's Instinct'' might be embedded close to those movies 
in the embedding space.
Movies with similar themes or genres (e.g., ``The Lord of the Rings'' (The LFR) and ``The Hobbit'') might also be embedded close together.
  
By finding similar movies and users in the embedding space, it is possible to recommend movies to users based on their preferences and according to similar movies in the embedding space.
From this embedding representation it can help to understand better relationships between different nodes in a graph, and its similarities.

A N-dimensional embedding represents each data point as a vector with N numerical values. 
These values, or coordinates, define the point's position in a N-dimensional space.
In general a method to create a N-dimensional embedding depends on the nature of your data and the task at hand.  
The choice of embedding technique and the specific methods used to create the numerical values will influence the properties of the embedding, 
such as its ability to preserve local structure, global structure, or other relevant characteristics~\cite{GOYAL201878}. 
\\
\begin{figure}[htbp]
    \centering
    \includegraphics[width=1.0\textwidth]{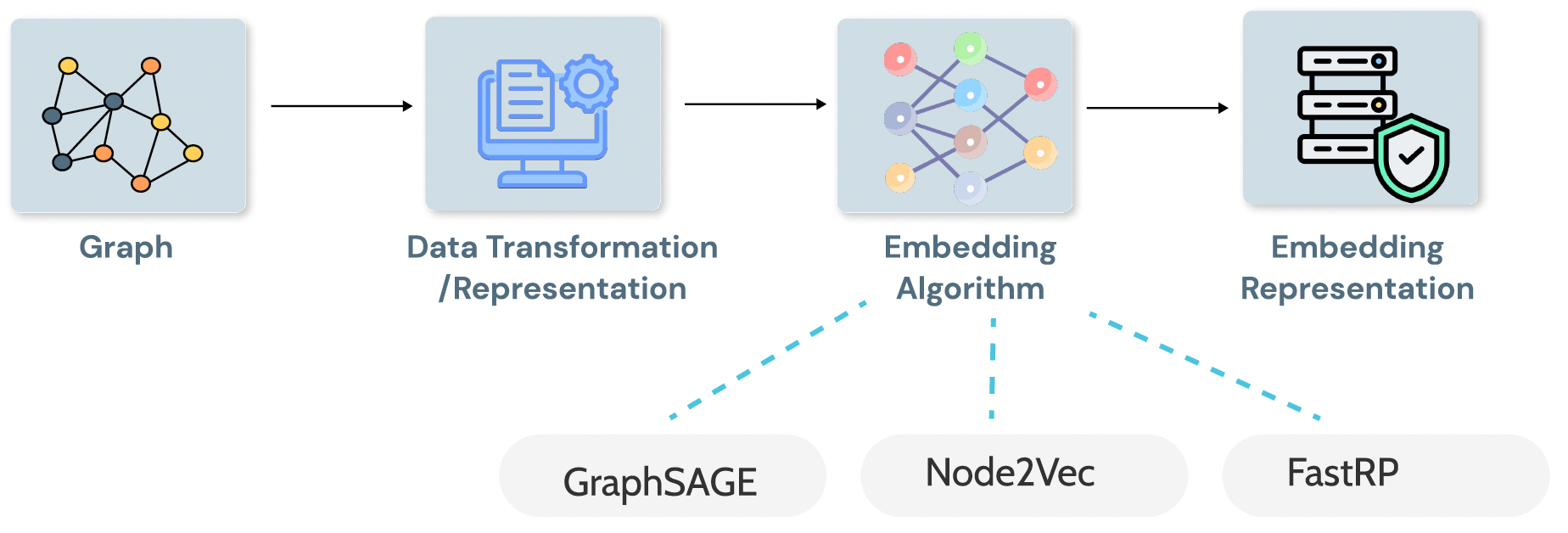}
    \caption{From Graph to Embedding}
    \label{fig:embedding} 
  \end{figure}

  As depicted in Figure~\ref{fig:embedding}, we can generate a vector representation, or embedding, for each node within the graph. 
  The specific embedding technique used, such as GraphSAGE or Node2Vec, will influence the characteristics of these vectors. 
  By analyzing these embeddings, we can extract important insights for a range of applications. 

Graph embedding, while powerful, presents several challenges. Heterogeneity arises when dealing with graphs containing diverse node and edge types,
necessitating specialized techniques. Scalability is a concern for large graphs, as many embedding methods can be computationally expensive. 
Moreover, interpretability can be a hurdle, as understanding the meaning of learned embeddings can be difficult for data analysts. 
Addressing these challenges is essential for effectively applying graph embedding to real-world problems.   

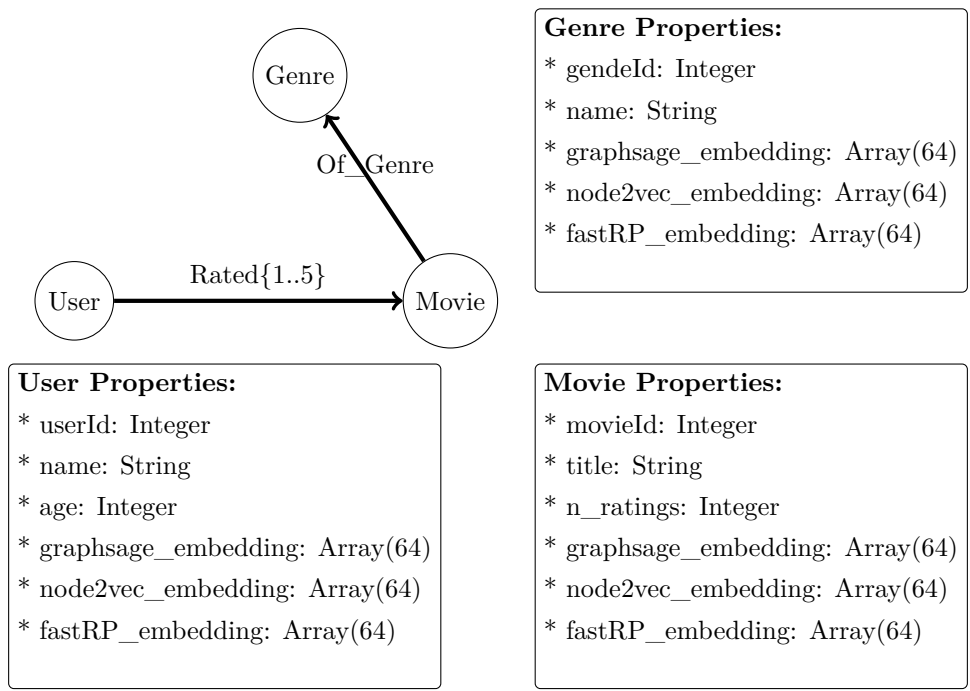
\begin{figure} 
    \centering
    \begin{tikzpicture}
        \node[circle,draw] (User) at (1,5) {User};  
        \node[circle,draw] (Movie) at (6,5) {Movie};
        \node[circle,draw] (Genre) at (4,8) {Genre};        
      
        \draw[->,ultra thick] (User) -- (Movie) node[midway,above] {Rated\{1..5\}};
        \draw[->,ultra thick] (Movie) -- (Genre) node[midway,above] {Of\_Genre};

        \node[draw, rounded corners=2pt, align=left] at (10,7) {
            \textbf{Genre Properties:}\\
            * gendeId: Integer \\
            * name: String\\             
            * graphsage\_embedding: Array(64)\\
            * node2vec\_embedding: Array(64)\\
            * fastRP\_embedding: Array(64)\\
          };
 
        \node[draw, rounded corners=2pt, align=left] at (3,2) {
            \textbf{User Properties:}\\
            * userId: Integer \\
            * name: String\\ 
            * age: Integer\\            
            * graphsage\_embedding: Array(64)\\
            * node2vec\_embedding: Array(64)\\
            * fastRP\_embedding: Array(64)\\
          };

        \node[draw, rounded corners=2pt, align=left] at (10,2) {
            \textbf{Movie Properties:}\\
            * movieId: Integer \\
            * title: String\\                       
            * n\_ratings: Integer\\            
            * graphsage\_embedding: Array(64)\\
            * node2vec\_embedding: Array(64)\\
            * fastRP\_embedding: Array(64)\\
          };

      \end{tikzpicture}
      \caption{Schema of \textit{MovieLens} Graph. \label{fig:schema}}
    \end{figure}

Figure~\ref{fig:schema} illustrates the graph schema, featuring nodes for Movies and Users (as depicted in Figure~\ref{fig:movie_graph}), while
 the 'Rated' edge connects these nodes, indicating user-movie interactions. Each node maintains a list of properties, including embeddings generated by using different approaches. 
 Storing embeddings as properties allows flexible analysis of different graph representations to determine which best suits specific tasks or real-world application. 
For instance, classification or recommendation tasks might favor one embedding approach over another.

    \begin{table}[h]
        \centering
        \begin{tabular}{ccccc}
        \toprule
        \textbf{Node Type} & \textbf{Node Name} & \textbf{graphsage\_emb} & \textbf{node2vec\_emb} & \textbf{fastRP\_emb} \\
        \midrule
        \rowcolor{lightgray}
        User & Bob  & [0.0025,...,0.0015]  & [0.0592,...,0.4480] & [0.0030,...,0.023] \\
        User & Alice  & [0.0025,...,0.0254], & [0.0088,...,0.1324] & [0.0652,...,0.0817] \\
        \rowcolor{lightgray}
        User & James  & [0.0,...,0.020] & [0.0025,...,0.0015] & [0.105,...,0.0154] \\
        Movie & The Hobbit & [0.0373,...,0.0399] & [0.0025,..., 0.00158] & [0.0709,...,0.0592] \\
        \rowcolor{lightgray}
        Movie & The LFR & [0.0025,...,0.029] & [0.0373,..., 0.0165] & [0.0154,...,0.0102] \\
        Movie & Anatomy & [0.027,...,0.0354] & [0.0592,..., 0.0940] & [0.0169,...,0.1323] \\
        \rowcolor{lightgray}
        Movie & MotherInstinct & [0.0025,...,0.0208] & [0.0399,..., 0.0354] & [0.0154,...,0.0687] \\
        \bottomrule 
        \end{tabular} 
        \caption{MovieLens Node Embeddings Example.} 
        \label{tab:embedding_example}
        \end{table} 

        Storing graph embeddings as node properties provides a flexible approach. 
        This allows for dynamic updates to embeddings as the graph changes and facilitates experimentation with different embedding techniques without 
        modifying the underlying graph structure. By comparing multiple embeddings, it is possible to evaluate their effectiveness for different tasks. 
        For instance, a deep learning-based node embedding might outperform a DeepWalk approach in specific classification or recommendation scenarios.  

        Table~\ref{tab:embedding_example} presents 
        sample node embeddings for the \textit{MovieLens} graph. Each node, representing either a User or a Movie, 
        is associated with three sets of embeddings generated using distinct techniques: GraphSAGE, Node2Vec, and FastRP. 
        These embeddings are conveniently stored as node properties, facilitating easy access and comparison. 
        By examining these embeddings, one can derive valuable insights into the relationships between nodes and their 
        similarities within the embedding space. 
        This information can be leveraged for a variety of applications, including recommendation systems, 
        community detection, and link prediction.

        \section*{Conclusion} 

        Graph embedding is a powerful technique for representing complex graph data in a lower-dimensional space.
        By transforming graphs into vectors, it simplifies the analysis and processing of large-scale datasets.
        Several approaches, such as GraphSAGE, Node2Vec, and FastRP, offer efficient methods for generating graph embeddings.
        By storing embeddings as node properties, it is possible to compare different embedding techniques and evaluate their effectiveness for specific tasks.        This flexibility allows for dynamic updates to embeddings and facilitates experimentation with different approaches.
        By analyzing these embeddings, one can extract valuable insights into the relationships between nodes and their similarities within the embedding space.
        This information can be leveraged for a range of applications, including recommendation systems, community detection, and link prediction.
        Addressing challenges such as heterogeneity, scalability, and interpretability is essential for effectively applying graph embedding to real-world problems.

\chapter{Analyzing Node Similarity from Graph Embedding Models}
\label{chap:similarity_embedding}

In this chapter, the main objective is to analyze the similarity between nodes from graph embedding. 
From embeddings it is possible to calculate embedding similarity and visualize the results. 
Compare the similarity results in the graph is a challenging, as it will be presented follow.
These results can be used to answer queries about the data, and to predict new relationships between nodes.
Visualize these similarities can also help to understand the dataset and also to improve some prediction results, by tunning some parameters/metrics. 
The similarities can be calculated by using different similarity metrics, such as cosine similarity, euclidean distance, and others.
Even if there are different ways to calculate the similarity between nodes, we have also different embedding methods to consider.
So, choose the most representative embedding method, the embedding dimension, 
the method to reduce the embedding dimensionality and the similarity metric is a big challenge.
 So, the main challenge to address is to choose the better embedding representation to an application problem. 
 Analyze the impact of these parameters on the similarity results is an open topic this chapter aims to address.

Consider the example where there is a graph with movies and genres as presented in the schema of 
Figure~\ref{fig:schema}. From this schema, consider only two types of nodes: 
Movie and Genre. The movies are connected to genres by the relationship \textit{OF\_GENRE}. 
The graph (\textit{Movielens dataset}) is represented in the Neo4J database and the embeddings are generated 
by Node2Vec and GraphSAGE methods. The embeddings are stored in the database and can be retrieved by 
Cypher~\cite{FrancisGGLLMPRS18} queries. The embeddings are used to verify the similarity between nodes and to visualize the results. 
To visualize the result about note similarity it is necessary to apply embedding reduce dimension methods\footnote{T-SNE is a method to visualize high-dimensional data by giving each data 
point a location in a two or three-dimensional map. The method is based on the probability distribution of the data points in the high-dimensional 
space and the map. The method tries to minimize the Kullback-Leibler divergence~\cite{Fernando08} between the two distributions. Isomap is a method 
to reduce the dimensionality of the data by preserving the geodesic distances~\cite{ShamaiK16} between all pairs of data points. The method is based on 
the graph representation of the data points and the shortest path between them. Spectral Embedding is a method to reduce the dimensionality of the data by preserving the pairwise distances between all pairs of data points. 
The method is based on the Laplacian matrix \cite{merris94} of the data points. MDS is a method to reduce the dimensionality of the data by preserving the pairwise distances between all pairs of data points.}
and it is done by t-SNE~\cite{Laurens12, JMLR:v15:vandermaaten14a, JMLR:v9:vandermaaten08a}, 
Isomap~\cite{tenenbaum_global_2000}, Spectral Embedding~\cite{Belkin:2003}, and MDS - Multidimention Scaling~\cite{Kruskal:1964io, Kruskal1964b} methods.

Consider the queries in Cypher~\cite{FrancisGGLLMPRS18} language to retrieve the embeddings for movies and its specific genres. The Cypher query for Node2Vec embedding is presented in Listing~\ref{lst:cypher-node2vec} 
and for GraphSAGE embedding is presented in Listing~\ref{lst:cypher-graphsage}. The embeddings are retrieved for movies that are connected to genres \textit{War} and \textit{Comedy}.

\begin{lstlisting}[language=cypher, numbers=left, label=lst:cypher-node2vec, caption=Cypher Query for Movie and Genre (Node2Vec Embedding)]
    MATCH (m:MOVIE)-[:OF_GENRE]->(genre:GENRE)       
    WHERE genre.name IN ["War", "Comedy"]
    RETURN m.name AS title, m.node2vec_embedding AS embedding, genre.name AS genre
\end{lstlisting}  

\begin{lstlisting}[language=cypher, numbers=left, label=lst:cypher-graphsage, caption=Cypher Query for Movie and Genre (GraphSAGE Embedding)]
    MATCH (m:MOVIE)-[:OF_GENRE]->(genre:GENRE)       
    WHERE genre.name IN ["War", "Comedy"]
    RETURN m.name AS title, m.graphsage_embedding AS embedding, genre.name AS genre
\end{lstlisting}

\begin{figure*}[hpbt]
    \centering
     \subfigure[]{%
        \includegraphics[width=0.45\linewidth]{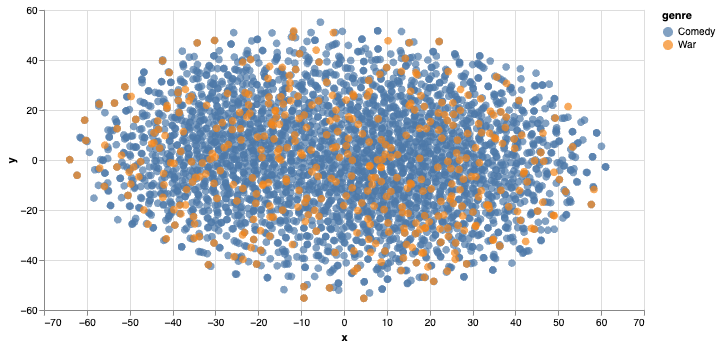}
        \label{fig:tSNE_node2vec}}
    \subfigure[]{
        \includegraphics[width=0.45\linewidth]{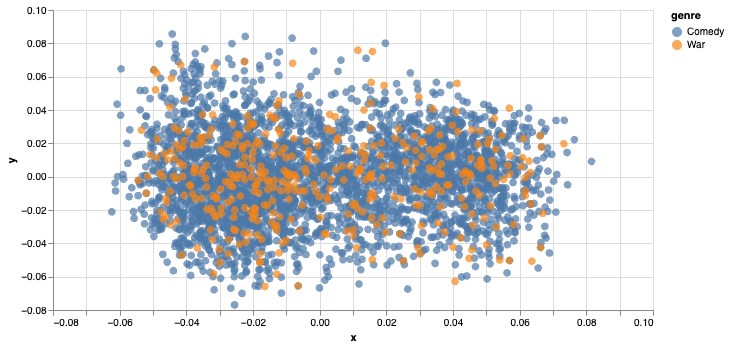}
        \label{fig:isomep_node2vec}}
    \subfigure[]{
        \includegraphics[width=0.45\linewidth]{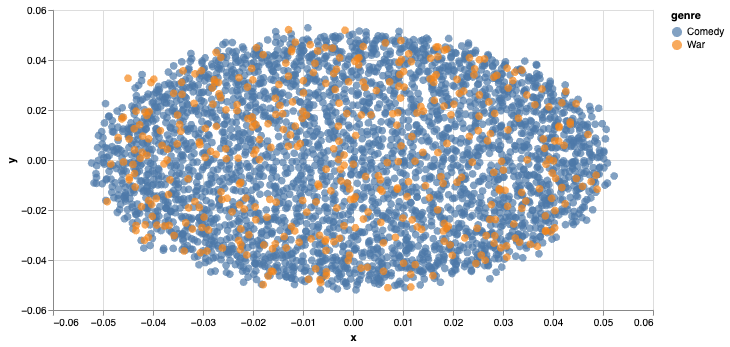}
        \label{fig:mds_node2vec}}        
    \subfigure[]{
        \includegraphics[width=0.45\linewidth]{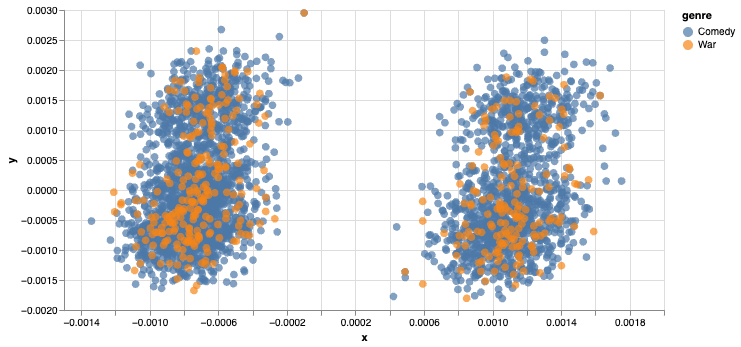}
        \label{fig:spectral_node2vec}} 
    
    \caption{TSNE, ISOMAP, MDS and Spectral Reduction with Node2Vec Embedding}\label{fig:reduction_node2vec}
\end{figure*}

\begin{figure*}[hpbt]
    \centering
     \subfigure[]{%
        \includegraphics[width=0.45\linewidth]{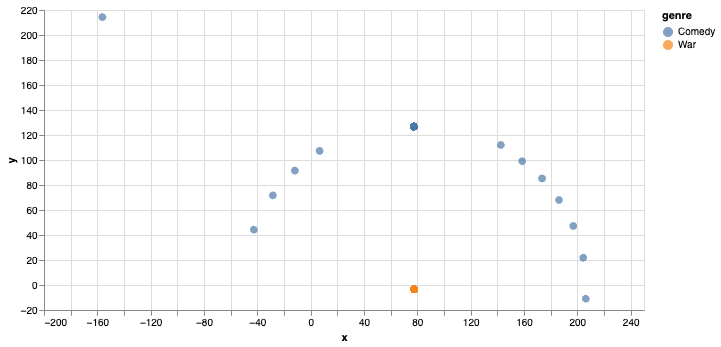}
        \label{fig:tsne_sage}}
    \subfigure[]{
        \includegraphics[width=0.45\linewidth]{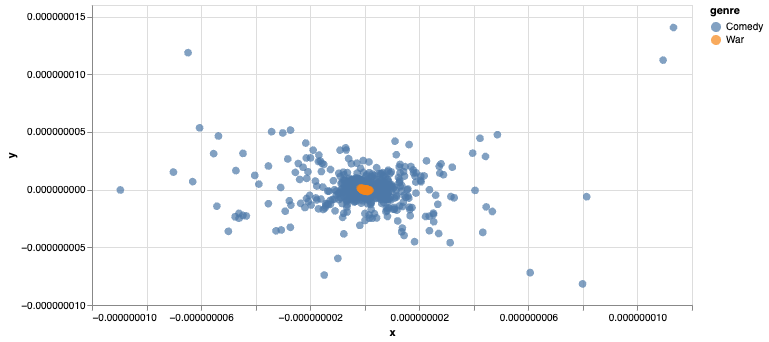}
        \label{fig:isomep_sage}}
    \subfigure[]{
        \includegraphics[width=0.45\linewidth]{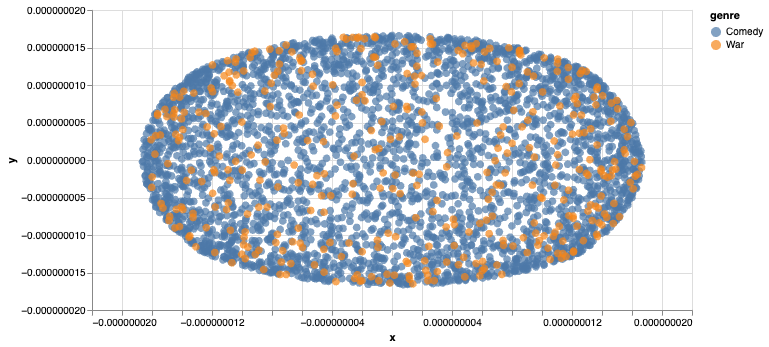}
        \label{fig:mds_sage}}        
    \subfigure[]{
        \includegraphics[width=0.45\linewidth]{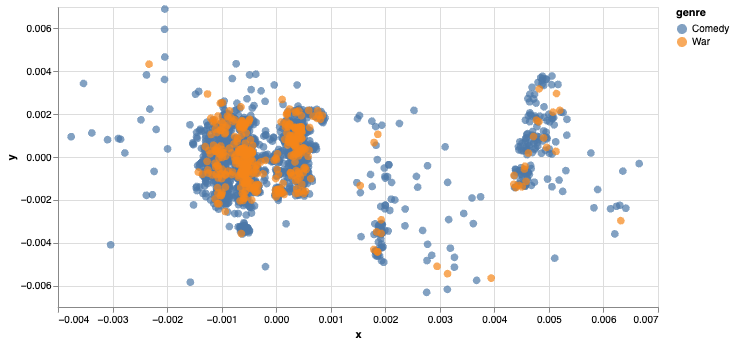}
        \label{fig:spectral_sage}} 
    
    \caption{TSNE, ISOMAP, MDS and Spectral Reduction with GraphSAGE Embedding}\label{fig:reduction_sage}
\end{figure*}

Figures~\ref{fig:reduction_node2vec} and~\ref{fig:reduction_sage} show the visualization of the embeddings for Node2Vec and GraphSAGE methods, respectively. 
The visualization is done by t-SNE (SubFigures \ref{fig:tSNE_node2vec} and \ref{fig:tsne_sage}), Isomap (SubFigures \ref{fig:isomep_node2vec} and \ref{fig:isomep_sage}), MDS (SubFigures \ref{fig:mds_node2vec} and \ref{fig:mds_sage}) and Spectral Embedding ((SubFigures \ref{fig:spectral_node2vec} and \ref{fig:spectral_sage})) methods. The visualization shows the similarity between nodes and the relationship 
between them. More close nodes are more similar than the distant ones. The visualization can help to understand the data and may also help to propose possible prediction results. 
Figures~\ref{fig:reduction_node2vec} and~\ref{fig:reduction_sage} are only a simple example of how to 
visualize the embeddings and the similarity between nodes. The visualization can be used to answer queries about the data and to infere possible new relationships between nodes. 
However, choose the better parameters and tunning them to generate the embeddings can impact the result. 

As it is possible to see about the Cypher queries and the latent space presented as the result, it is necessary to understand the problem and the data to 
choose the better embedding method. Maybe the Node2Vec method is better than GraphSAGE for a specific problem, or vice-versa.

  Imagine a Query $Q$ that wants to know about the similarity between the movies \textit{The Hobbit} and other movies in the dataset, or 
  a Query $S$ that wants to know possible movies to suggest to a user that likes \textit{Anatomy of a Fall}, such as Alice or James (Figure \ref{fig:movie_graph}). 
  Considering the embeddings presented, maybe the results could be different if the embeddings were generated by a method A or a method B, or if the dimensionality is different.
  The Query $Q$ can be written as $Q = \text{similarity}(\text{The Hobbit}, m: Movie \in MOVIES)$ 
  and the Query $S$ as $S = \text{similarity}(\text{Anatomy of a Fall}, \text{Alice})$ .

  Query $Q$ can be answered by calculating the similarity between the embeddings of the movies \textit{The Hobbit} and other movies in the dataset, 
  while Query $S$ can be answered by calculating the similarity between the embeddings of the movie \textit{Anatomy of a Fall} and the user Alice.    
  But it is not so simple to calculate the similarity between the embeddings. There are different ways to calculate the similarity between nodes, such as cosine similarity, euclidean distance, and others.
    Even if there are different ways to calculate the similarity between nodes, we have also different embedding methods that can generate different results.
    The challenge, as we already mentioned, is to choose the most representative embedding method, the embedding dimension, the method to reduce this dimensionality and the similarity metric to use the best embedding representation to an application problem.
    So, choose the better parameters and tunning them to generate the embeddings can impact the result. 
    We advocate that the use of specific embedding configuration to type of problems (Queries) can improve the accuracy of the results.
    This rises the question: How to choose the better embedding representation to an application problem? We believe that group of queries can be used to choose/define the better embedding representation to answer application needs.

    In Neo4J database it is possible to use the Graph Data Science (GDS)~\cite{HodlerN22} library to calculate the similarity between nodes. KNN algorithm is a native function to verify a score over a set of similar embeddings.
    The Cypher query to calculate the similarity between nodes using the KNN algorithm~\cite{Kramer2013} is presented in Listing~\ref{lst:cypher-knn}. The KNN algorithm calculates the similarity between nodes and stores the results in the database.
    This function generate a new \textit{Edge} named \textit{SIMILAR} between movie nodes and stores the similarity score as a edge property named \textit{score}. In this example the topK is 5, the nodeProperties is the embedding property,
    the randomSeed is 42, the deltaThreshold is 0.7, the writeRelationshipType is \textit{SIMILAR}, and the writeProperty is \textit{score}. Once we have new edges the represents similarity in the graph, 
    it is possible to use the Cypher query to retrieve new information. In the example presented in Listing~\ref{lst:cypher-knn} it considered the \textit{graphsage\_embedding} stored in each movie node for the KNN input.
    As well as it is possible to generate new relationships between movie nodes, it is possible to do the same for user nodes. 
    User are similar if they have similar ratings for the same movies. 
    \begin{lstlisting}[language=cypher, numbers=left, label=lst:cypher-knn, caption=KNN over Graph Embeddings]
        CALL gds.knn.write('RatingKNN', {
            topK: 5,
            nodeProperties: ['graphsage_embedding'],
            randomSeed: 42,        
            deltaThreshold: 0.7, 
            writeRelationshipType: "SIMILAR",
            writeProperty: "score"
        })
        YIELD nodesCompared, relationshipsWritten, similarityDistribution
        RETURN nodesCompared, relationshipsWritten, similarityDistribution.mean as meanSimilarity
\end{lstlisting}  

The Cypher query to predict possible user ratings for movies is presented in Listing~\ref{lst:cypher-rating}. 
The query considers the user nodes and the movie nodes listed as the 5 ids for users (\textit{uids}) the most rated movies, and the 5 ids for movies (\textit{mids}) the most rated.
\textit{UNWIND} is used to iterate over the lists of user ids and movie ids.
The queries tries to predict a possible rating for the movies that the user has already rated, considering their similar. 
The query compares the rating for the 5 user's similar and the rating for the user.
Doing this, it would be possible to predict eventual rating for movies that the user has not rated yet.
The same can be done for users, considering the movies that they have rated, by suggesting new movies to watch by analyzing rating averages for similar movie.
Line 5 retrieves the user nodes and the movie nodes that the user has rated. Line 6 retrieves the user nodes and the movie nodes that the user has not rated yet, but other similar users have rated.
Line 7 returns the user id, the movie id, the predicted rating, and the real rating. 
The query can be used to predict possible user ratings for movies and to suggest new movies to users.

    \begin{lstlisting}[language=cypher, numbers=left, label=lst:cypher-rating, caption=Predict Possible User Rating for Movies]
        UNWIND ['574', '564', '624', '15', '73'] AS uids 
        UNWIND ['356', '296', '318', '593', '260'] AS mids
        CALL{
            WITH uids, mids
            MATCH (u1:User {userId:uids})-[r1:RATED]->(m1:Movie{movieId:mids})
            MATCH (u:User {userId:uids})-[s:SIMILAR]->(:User)-[r:RATED]->(m:Movie{movieId:mids}) 
            RETURN u.userId AS userId, m.movieId as movie, avg(r.rating) as prediction_rating, r1.rating as real_rating     
        } return userId, movie, prediction_rating, real_rating
    \end{lstlisting}

    \begin{table}[htbp]
        \centering
        \begin{tabular}{ccccccc}
        \toprule
        \textbf{userId} & \textbf{movie} & \textbf{title} & \textbf{prediction\_rating} & \textbf{real\_rating}  & \textbf{difference} \\
        \midrule
        \rowcolor{lightgray}
        574 & 356  & Forrest Gump  & 4.13	& 4.50& 	0.38 \\
        574 & 296  & Pulp Fiction & 4.13	& 5.00	& 0.88\\
        \rowcolor{lightgray}
        574 & 318  & The hawshank Redemption  & 5.00	& 5.00	& 0.00\\
        574 & 260  & Star Wars: Episode IV & 4.33	& 4.00	& -0.33\\
        \rowcolor{lightgray}
        564 & 356  & Forrest Gump  & 3.75& 	3.00	& -0.75\\
        564 & 296  & Pulp Fiction & 4.50	& 5.00& 	0.50\\
        \rowcolor{lightgray}
        564 & 593  & The Silence of the Lambs  & 4.13	& 5.00	& 0.88 \\
        564 & 260  & Star Wars: Episode IV  & 4.38	& 2.00& 	-2.38\\
        \rowcolor{lightgray}
        624 & 356  & Forrest Gump  & 3.67	& 3.00& 	-0.67\\
        624 & 296  & Pulp Fiction & 4.50 & 	5.00	& 0.50\\
        \rowcolor{lightgray}
        624 & 593  & The Silence of the Lambs  & 4.33	& 5.00 & 	0.67\\
        624 & 260  & Star Wars: Episode IV & 4.63	& 5.00	& 0.38\\
        \rowcolor{lightgray}
        15 & 356  & Forrest Gump & 5.00	& 1.00	& -4.00\\
        15 & 318  & The Shawshank Redemption & 3.00	& 2.00	& -1.00\\
        \rowcolor{lightgray}
        15 & 593  & The Silence of the Lambs  & 5.00	& 5.00	& 0.00 \\
        15 & 260  & Star Wars: Episode IV  & 2.67	& 5.00	& 2.33\\
        \rowcolor{lightgray}
        73 & 356  & Forrest Gump  & 5.00	& 5.00	& 0.00 \\
        73 & 296  & Pulp Fiction & 4.50	& 5.00	& 0.50\\
        \rowcolor{lightgray}
        73 & 318  & The Shawshank Redemption  & 5.00	& 5.00	& 0.00 \\
        73 & 593  & The Silence of the Lambs & 3.50	& 4.50	& 1.00\\
        \rowcolor{lightgray}
        73 & 260  & Star Wars: Episode IV  & 4.50	& 4.50 &	0.00 \\        
        \bottomrule 
        \end{tabular} 
        \caption{Rating Prediction Using GraphSAGE Embeddings.} 
        \label{tab:prediction_graphSage}
        \end{table}

        \begin{table}[htbp]
            \centering
            \begin{tabular}{ccccccc}
            \toprule
            \textbf{userId} & \textbf{movie} & \textbf{title} & \textbf{prediction\_rating} & \textbf{real\_rating}  & \textbf{difference} \\
            \midrule
            \rowcolor{lightgray}
            574 & 356  & Forrest Gump  & 4.60 & 	4.50	& -0.10 \\
            574 & 296  & Pulp Fiction & 5.00	& 5.00	& 0.00\\
            \rowcolor{lightgray}
            574 & 318  & The hawshank Redemption  & 4.75	& 5.00	& 0.25\\
            574 & 260  & Star Wars: Episode IV & 5.00	& 4.00	& -1.00\\
            \rowcolor{lightgray}
            564 & 356  & Forrest Gump  & 3.00	& 3.00	& 0.00\\
            564 & 296  & Pulp Fiction & 5.00	& 5.00	& 0.00\\
            \rowcolor{lightgray}
            564 & 593  & The Silence of the Lambs  &4.00	& 5.00	& 1.00\\
            564 & 260  & Star Wars: Episode IV  &4.50	& 2.00	& -2.50\\
            \rowcolor{lightgray}
            624 & 356  & Forrest Gump  & 3.70& 	3.00	& -0.70\\
            624 & 296  & Pulp Fiction & 3.75	& 5.00& 	1.25\\
            \rowcolor{lightgray}
            624 & 593  & The Silence of the Lambs  &3.64	& 5.00& 	1.37\\
            624 & 260  & Star Wars: Episode IV & 4.50	& 5.00	& 0.50\\
            \rowcolor{lightgray}
            15 & 356  & Forrest Gump & 4.50	& 1.00	& -3.50\\
            15 & 318  & The Shawshank Redemption & 4.60	& 2.00	& -2.60\\
            \rowcolor{lightgray}
            15 & 593  & The Silence of the Lambs  & 3.60	& 5.00	& 1.40\\
            15 & 260  & Star Wars: Episode IV  & 3.50	& 5.00	& 1.50\\
            \rowcolor{lightgray}
            73 & 356  & Forrest Gump  & 3.50	& 5.00	& 1.50\\
            73 & 296  & Pulp Fiction & 5.00& 	5.00	& 0.00\\
            \rowcolor{lightgray}
            73 & 318  & The Shawshank Redemption  & 4.00	& 5.00	& 1.00 \\
            73 & 593  & The Silence of the Lambs & 4.00	& 4.50	& 0.50\\
            \rowcolor{lightgray}
            73 & 260  & Star Wars: Episode IV  & 4.50	& 4.50	& 0.00\\        
            \bottomrule 
            \end{tabular} 
            \caption{Rating Prediction Using Node2Vec Embeddings.} 
            \label{tab:prediction_node2vec}
            \end{table}

Tables \ref{tab:prediction_graphSage} and \ref{tab:prediction_node2vec} show the results from Listening \ref{lst:cypher-rating} for the users 574, 564, 624, 15, and 73.
Tables \ref{tab:prediction_graphSage} and \ref{tab:prediction_node2vec} tries to analyse some kind of rating prediction for the movies \textit{Forrest Gump}, 
\textit{Pulp Fiction}, \textit{The Shawshank Redemption}, \textit{Star Wars: Episode IV}, and \textit{The Silence of the Lambs} for the users 574, 564, 624, 15, and 73.
The prediction results are calculated by using the GraphSAGE and Node2Vec embeddings. The prediction results are compared to the real ratings.
The difference between the prediction rating and the real rating is presented in the column \textit{difference}.

For both analyses, the GraphSAGE embeddings presented better results than the Node2Vec embeddings. The difference between the prediction rating and the real rating is smaller for the GraphSAGE embeddings than for the Node2Vec embeddings.
Even if both embeddings predict correctly 5 movies, the difference between the prediction rating and the real rating is smaller for the GraphSAGE embeddings than for the Node2Vec embeddings.
Only 5 prediction rating have a different grater or equal to 1 for the GraphSAGE embeddings, while 12 predictions have a different grater or equal to 1 for the Node2Vec embeddings.
The GraphSAGE embeddings presented a better prediction rating than the Node2Vec embeddings. The GraphSAGE embeddings are more accurate than the Node2Vec embeddings for this simple example of predicting rating.

\section*{Conclusion}

In this chapter, we presented a simple way to analyze the similarity between nodes from graph embeddings.
The example used considered embeddings generated by Node2Vec and GraphSAGE methods and stored in the Neo4J database.
The embeddings were retrieved by Cypher queries and the similarity between nodes was calculated by cosine similarity.
The similarity results were visualized by t-SNE, Isomap, Spectral Embedding, and MDS methods.
The visualization showed the similarity between nodes and the relationship between them.
It is possible to see that node similarities may improve accuracy about queries results.
The visualization can help to understand the data and may improve possible prediction results.
The impact of these parameters on the similarity results is an interesting open topic.

So, from these general discussion about embedding, similarities and graph queries, some questions can be raised:

\begin{itemize}
    \item Is it possible have similar results from different embedding representation built from the same graph or subgraph data? 
    \item Can we have a Predictive Query Language for Graph applying a set o techniques in order to find the better result for the user need?
\end{itemize}

\chapter{Predictive Query-based Pipeline}
\label{chap:predictive_query}

Predictive queries over database system it is not a easy task. There are some initiatives to merge artificial intelligence 
and relational database domains\cite{Cyril05, 0006TXLHQZ0ZL24, FouladvandGNSNJ23} in order to reach predictive important queries over the dataset. 
Graph database has a important role in data representation and analysis. In the last years different works \cite{0002YRWL23,Cyril05} 
consider graph schemas for data representation and node or edge prediction and classification. 

However, it is a expensive task to predict information over graphs, specially if the user/expert does not know what type 
of queries solve her problem. Two important questions to be considered can be: \textit{(i) How much a given query Q is adapted to a given graph database G?} and
\textit{(ii) How can we evaluate whether Q ``makes sense''/``is adapted'' to G?}. From these 2 questions, one of our main objectives is to define a quality indice of the answers for query Q on G (without executing Q on G).
For instance, if we want to predict the grade given to movies, but there are just few films having a grade, the quality of the answers will not be good.

Nowadays there are several libraries that provide implementations of graph embedding algorithms, and 
some of them are designed to work with machine learning libraries, such as TensorFlow~\cite{AbadiBCCDDDGIIK16} and PyTorch~\cite{PaszkeGMLBCKLGA19}.
For example, the library \textit{GraphSAGE}~\cite{DBLP:conf/nips/HamiltonYL17} is implemented in TensorFlow, 
and it is designed to work with the machine learning library \textit{scikit-learn}~\cite{PedregosaVGMTGBPWDVPCBPD11}.
NetworkX~\cite{Hagberg08} is a Python package for creating and analyzing graphs, with support for graph embedding algorithms.
StellarGraph~\cite{StellarGraph} is also a Python library for graph machine learning, providing a high-level API for graph embedding and prediction tasks.
And the library \textit{Deep Graph Library}~\cite{abs-1909-01315} (DGL) is a Python package for building graph neural networks, with support for graph embedding algorithms.
There are many other libraries that provide implementations of graph embedding algorithms, 
and some of them are designed to work with machine learning libraries. The decision of which library to use depends on 
the specific requirements of the application, such as the size of the graph, the type of graph data, and the machine learning algorithms 
that will be used.  

Graph Neural Networks~\cite{ZhouCHZYLWLS20} (GNNs) are a class of neural networks specifically designed to process graph-structured data. 
They leverage the graph's structural information to learn meaningful representations of nodes and edges. 
GNNs are composed of multiple layers, each of which aggregates information from neighboring nodes and updates the node's representation.

Graph Convolucional Networks~\cite{LiHW18} (GCNs)are a popular type of GNN that extend convolucional neural networks to graph-structured data. 
They operate by convolving a filter over each node's neighborhood, capturing the structural information of the graph.

GCN applications can be: \textit{(i) Node classification}, predicting the labels of nodes in a graph;
\textit{(ii)) Link prediction}, predicting the existence or strength of edges between nodes; \textit{(iii) 
Community detection}, identifying groups of nodes that are densely connected and also
\textit{(iv) Graph clustering}, partitioning a graph into clusters of similar nodes.

In summary, GNNs and GCNs are powerful tools for processing graph-structured data and have a wide range of applications in various domains.
So, graph embedding is used as input data for GNNs and GCNs, and that run specific models to build machine learning models for graph data.
These models can be used to answer possible ``\textit{predicting queries}'' about the graph, such as predicting the labels of nodes, predicting the existence of edges, 
and identifying communities of nodes.

\begin{figure}[h]
  \centering
  \includegraphics[width=0.95\textwidth]{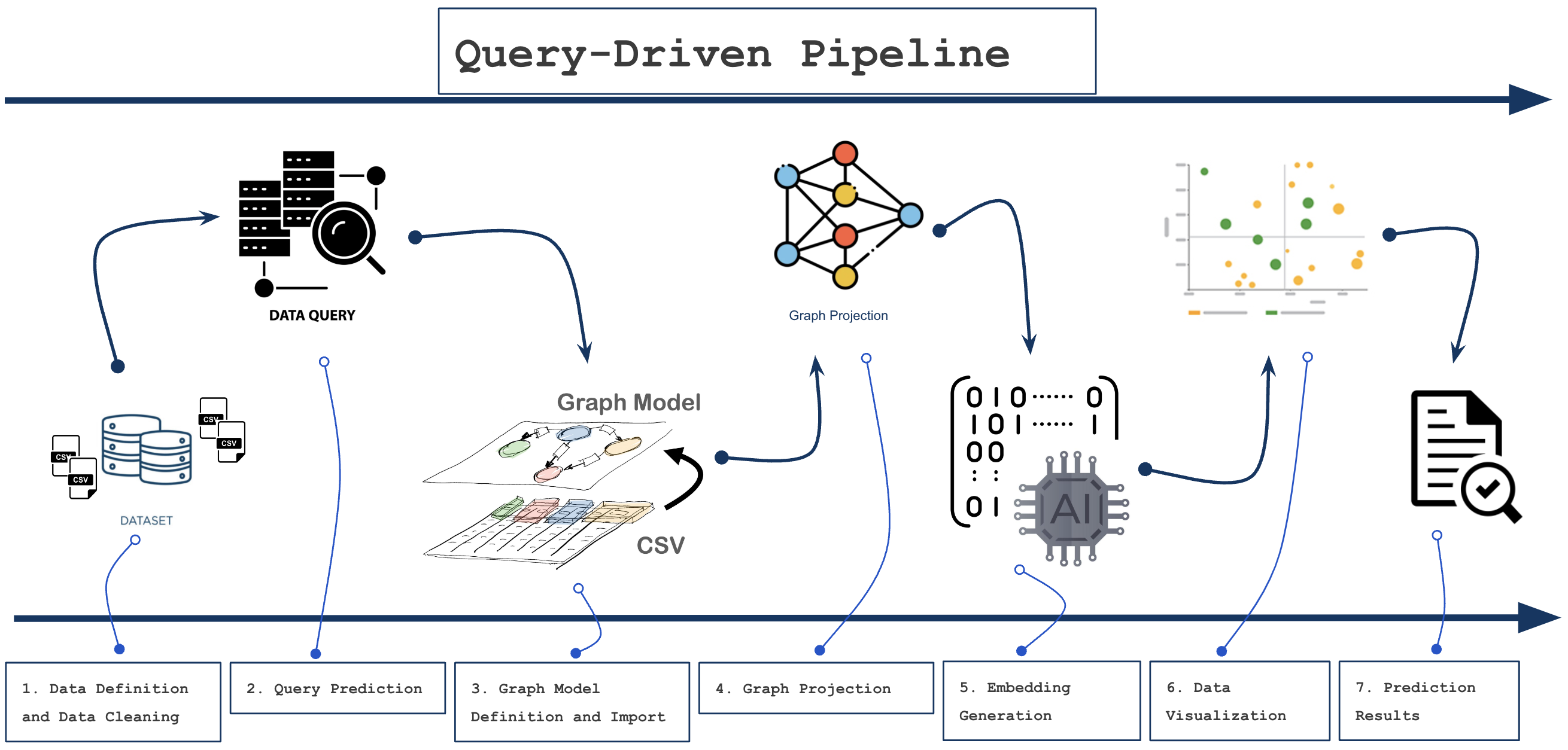}
  \caption{Query-Driven Pipeline}
  \label{fig:pipeline} 
\end{figure}

To enable predictive analysis based on specific data, we propose a pipeline that effectively transforms data in general 
formats into a graph representation. By employing embedding methods and specialized data projections, we can leverage 
artificial intelligence similarity algorithms to generate predictions on the original data, addressing relevant queries. 
Figure \ref{fig:pipeline} illustrates our proposed methodology (pipeline) for information prediction within data graphs.

The steps of the proposal pipeline (Figure \ref{fig:pipeline}) are the following:
\begin{enumerate}
  \item Data Definition and Data Cleaning;
  \item Query Prediction;
  \item Graph Model Definition and Import;
  \item Graph Projection;
  \item Embedding Generation;
  \item Data Visualization;
  \item Prediction Results.
\end{enumerate}

When working with a specific dataset, it's crucial to carefully select the most representative attributes 
and clean the data to remove incomplete or irrelevant information. This data preparation step is essential 
for effectively mapping the data to a graph database and defining the queries we intend to answer or the 
predictions we aim to make. Based on the selected dataset and its information content, the second step involves defining the specific 
queries we aim to answer, taking into account predictive aspects. Each query will be carefully analyzed in 
relation to the parameters of the subsequent stages in the pipeline. The third step entails importing the chosen 
dataset into a graph database, such as Neo4J. A well-defined schema that accurately represents our data and desired queries is essential. 
This schema will significantly influence the projection and generation of embeddings. 
As illustrated in Figure \ref{fig:pipeline}, the fourth step involves defining various projection types. 
The choice of projection method can have a substantial impact on the resulting responses from the graph.

 Thus, we defined 3 types of possible projections for data analysis. The objective of the pipeline is to explore important aspects considering a graph projection. We aim to observe the impact of projection \textit{Q}, \textit{i.e.} 
which is the better projection for à given query (\textit{n-neighborhood} projection). It can be a (i) \textit{full graph projection}, (ii) a \textit{strict graph projection} (2 nodes and 1 edge), and 
(iii) a n-neighborhood projection (more than 2 nodes, and more than 1 edge), that we named \textit{strict-extended graph projection}.

The pipeline also aims to analyze the impact of different embedding methods (taking into account topology and attributes) over the query result.
For this, it is consider 3 different types of methods for generating embeddings, they are: \textit{Node2Vec, GraphSAGE} and \textit{FastRP}. For each embedding generation, different
dimensions are considered. We propose the use of 10, 50 and 100 embedding dimensions, as shown in Table~\ref{table:project}.

\begin{table}[h!]
  \centering
  \begin{tabular}{||c c c||} 
   \hline
    Projection Types & Embedding Methods & Dimensions  \\ [0.5ex] 
   \hline\hline
     Full Graph Projection & Node2Vec & 10 \\
     Strict Graph Projection & GraphSAGE & 50 \\
     Strict-Extended Graph Projection & FastRP & 100 \\ [1ex] 
   \hline
  \end{tabular}
  \caption{Projection Types | Embedding Methods | Dimensions.}
  \label{table:project}
  \end{table}

  By the embedding generated from each projection/method and we reduce the embeddings to a latent space in order to visualize the data. 
  From the reduction we can search for similarities between the data in order to answers the original prediction queries made in step 2.
  For each embedding method, as shown in Table~\ref{table:project}, we will analyze not only the impact of the embedding dimension on the prediction results, 
  but also if the type of projection will impact the result.
  
For this, it is considered a public datasets used for the experiments. The dataset was collected from Kaggle \footnote{Here there are also other interesting datasets,
 like, Life Expectancy~\cite{DBK:Life} and Mental Health Prediction~\cite{DBK:Mental} that may be considered in future works. }, 
considering the chosen theme Heart Disease~\cite{DBK:Heart}. 
All data description and details about properties of this set of data can be found in the references~\cite{DBK:Heart}.

From the Health Dataset the queries~\footnote{It is important to highlight that these queries are examples non exhaustive to conduct the pipeline validation.} to conduct the pipeline execution is the follow.

\begin{itemize}
  \item Heart Disease Data:  
  \begin{itemize}
    \item \textit{Does Person X have heart disease?}    
    \item \textit{If you have Cholesterol and Sugar disorder, it implies directly in heart disease?} 
  \end{itemize}  
\end{itemize}

As previously discussed, the initial step in developing a machine learning model for graph data involves 
defining the embedding parameters that will serve as input for the neural network. 
This entails determining which nodes will be used as input data for the model and 
how the embeddings will be employed to represent these nodes within the graph. 
Additionally, it's essential to decide whether the entire graph or a specific subset (subgraph) 
will be considered. The dimensionality of the embeddings and the update mechanism during the training process
 must also be carefully specified.
 The appropriate dimensionality of the embeddings used as input data for the model can vary depending on the specific 
 application, ranging from 10 values to 1024 values. In the case of GCN models, 
 the embeddings can be dynamically updated during the training process, allowing for adjustments
  to the dimension vector. The embedding model itself will also play a role in the overall process.
   Consequently, several parameters need to be carefully defined before constructing a machine learning model for graph data.
   The health dataset we utilized contains information about patients and their associated heart medical conditions.
    The graph structure consists of nodes representing individual patients, connected by edges that 
    signify relationships between them. Each patient record within the dataset includes various properties, such as:

\begin{itemize}
    \item age
    \item sex
    \item chest pain type (4 values)
    \item resting blood pressure
    \item serum cholestoral in mg/dl
    \item fasting blood sugar \> 120 mg/dl
    \item resting electrocardiographic results (values 0,1,2)
    \item maximum heart rate achieved
    \item exercise induced angina
    \item oldpeak = ST depression induced by exercise relative to rest
    \item the slope of the peak exercise ST segment
    \item number of major vessels (0-3) colored by flourosopy
    \item thal: 0 = normal; 1 = fixed defect; 2 = reversable defect
\end{itemize}

The ``target'' field refers to the presence of heart disease in the patient. It is integer valued 0 = \textit{no disease} and 1 = \textit{disease}.

\begin{figure} 
    \centering
    \begin{tikzpicture}
        \node[circle,draw] (Person) at (0,0) {Person};  
        \node[circle,draw] (PersonState) at (-3,3) {PersonState};
        \node[circle,draw] (HeartMeasures) at (3,3) {HeartMeasures};    
        \node[circle,draw] (HeartÈxams) at (5,0) {HeartÈxams};        
        \node[circle,draw] (FS) at (0,-3) {FS};        
        \node[circle,draw] (DiseaseResult) at (-5,0) {DiseaseResult};        
      
        \draw[->,ultra thick] (Person) -- (FS) node[midway,above] {hasFS};
        \draw[->,ultra thick] (Person) -- (HeartMeasures) node[midway,above] {hasHeartMeasures};
        \draw[->,ultra thick] (Person) -- (PersonState) node[midway,above] {hasState};
        \draw[->,ultra thick] (Person) -- (HeartÈxams) node[midway,above] {hasHeartExams};
        \draw[->,ultra thick] (Person) -- (DiseaseResult) node[midway,above] {hasDisease};

    \end{tikzpicture}
    \caption{Schema of \textit{Heart-Desiase} Graph. \label{fig:heart-disease-graph}}
  \end{figure}
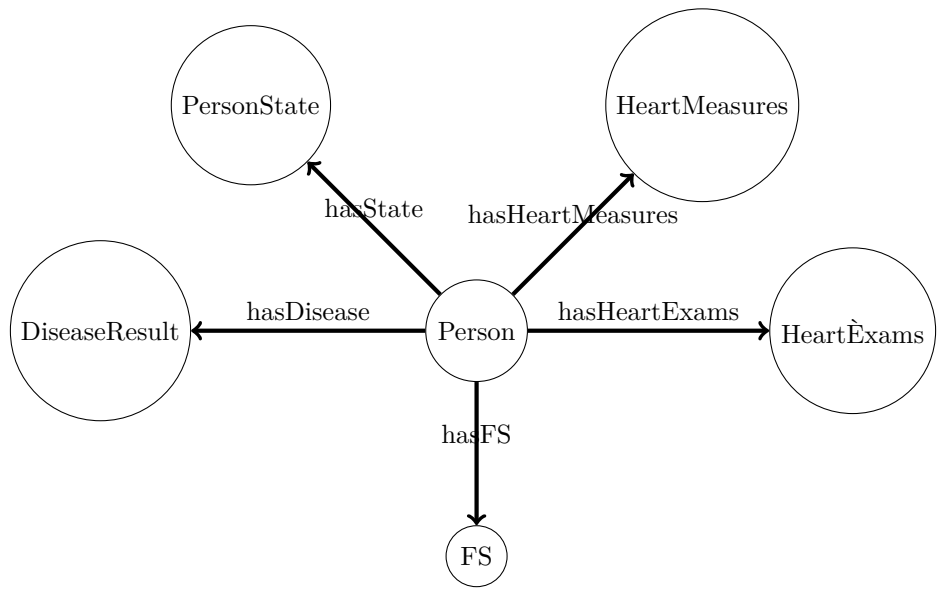

From the original dataset, a graph was created with the following properties:
\begin{itemize}
    \item Nodes:
    \begin{itemize}
        \item Person;
        \item PersonState;
        \item HeartMeasures;
        \item HeartÈxams;
        \item FS (Fat and Sugar);
        \item DiseaseResult;
    \end{itemize}
    \item Edges:
    \begin{itemize}
        \item hasFS;
        \item hasHeartMeasures;
        \item hasState;
        \item hasHeartExams;
        \item hasDisease;
    \end{itemize}
\end{itemize}

Figure \ref{fig:heart-disease-graph} shows the schema of the \textit{Heart-Desiase} graph. The graph has 933 nodes and 1818 edges.
Each node in the graph represents a patient or the heart information, and each edge represents a relationship between a person and his condition.
The goal is to build a machine learning model that can predict the presence of heart disease in a patient based on the patient's information and heart condition.   

\begin{lstlisting}[language=Python, caption=Cypher Code - Full Projection, label={lst:full-project}]
call gds.graph.project(
  'fullgraphsage',
  {
    Person: {
      properties: ['age', 'gender']
    },
    PersonState: {
      properties: ['cp', 'thal']
    },
    FS: {
      properties: ['type', 'value']
    },
    HeartExames: {
      properties: ['ca', 'exang', 'oldpeak', 'restecg', 'slope']
    },
    HeartMesures: {
      properties: ['thalach', 'trestbps']
    },
    DiseaseResult: {
      properties: ['target']
    },
  }, {
    hasDisease: {
      orientation: 'undirected'
    },
    hasFS: {
      orientation: 'undirected'
    },
    hasHeartExames: {
      orientation: 'undirected'
    },
    hasHeartMesures: {
      orientation: 'undirected'
    },
    hasState: {
      orientation: 'undirected'
    }
})
\end{lstlisting}

The Cypher code example in Listing~\ref{lst:full-project} demonstrates how to project the entire graph. 
The code defines the properties of each node and edge within the graph, as well as the relationships between them. 
The full graph projection is created using the \textit{gds.graph.project} procedure, which takes the graph name as input. 
Strict Graph Projection and Strict-Extended Graph Projection are generated in a similar manner, 
with the distinction lying in the different definitions of properties and relationships for each projection.
Strict Graph Projection limits the scope to a single node and onde directly connected edge. 
In contrast, Strict-Extended Graph Projection encompasses a broader set, 
including the node itself, its directly connected edges, and at least one additional node or edge. 

Embeddings were generated using the Node2Vec, GraphSAGE, and FastRP algorithms, 
with dimensions of 10, 50, and 100 for all projection types. This resulted in a total of nine distinct embeddings 
for the graph. These embeddings were subsequently employed as input data for the machine learning model, 
constituting a valuable contribution of this work. 
Other researchers can utilize this dataset of embeddings for the Heart Disease dataset to 
investigate the impact of embeddings on various types of machine learning models.

\begin{figure*}[hpbt]
  \centering
   \subfigure[]{%
      \includegraphics[width=0.45\linewidth]{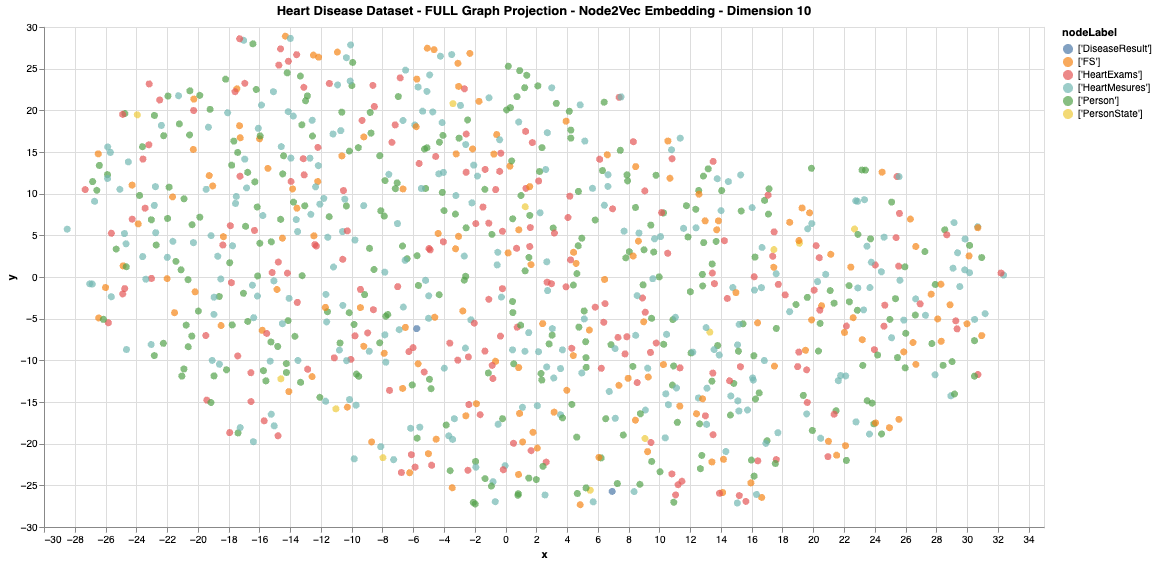}
      \label{fig:full_node2vec_10}}
  \subfigure[]{
      \includegraphics[width=0.45\linewidth]{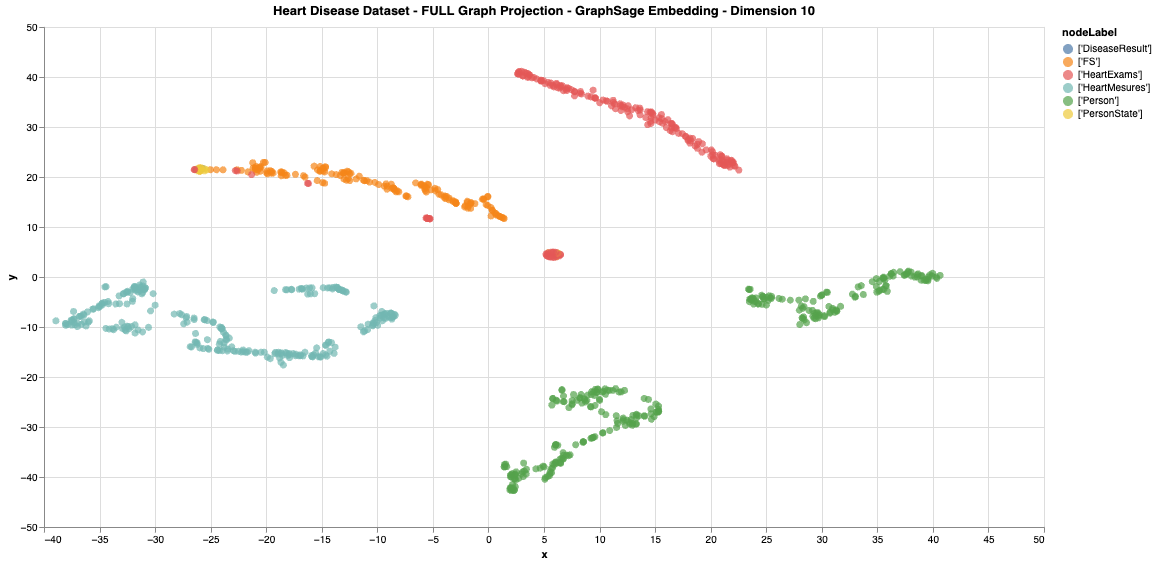}
      \label{fig:full_graph_sage_10}}
  
  \caption{Full Graph Projection - Node2Vec and GraphSAGE - Dimension 10}\label{fig:10dim}
\end{figure*}

Figures \ref{fig:full_node2vec_10} and \ref{fig:full_graph_sage_10} show the embeddings generated for the Full Graph Projection using the 
Node2Vec and GraphSAGE algorithms with 10 dimensions for all nodes. As the embeddings are in a latent space, and the nodes 
are represented as points in the space, each color is a type of node. Notice that in Figure \ref{fig:full_graph_sage_10}, 
the nodes are more separated in the space, which means that the embeddings are more discriminative for the GraphSAGE algorithm.
While in Figure \ref{fig:full_node2vec_10}, the nodes are more mixed in the space, which means that
the embeddings are less discriminative for the Node2Vec algorithm. The GraphSAGE embedding classifies 
the nodes better than the Node2Vec using a 10-dimension embedding. The is no information about the quality of the embeddings,
but it is possible to see that the GraphSAGE embeddings are, is this case, more representative for classification.

\begin{figure*}[hpbt]
  \centering
   \subfigure[]{%
      \includegraphics[width=0.45\linewidth]{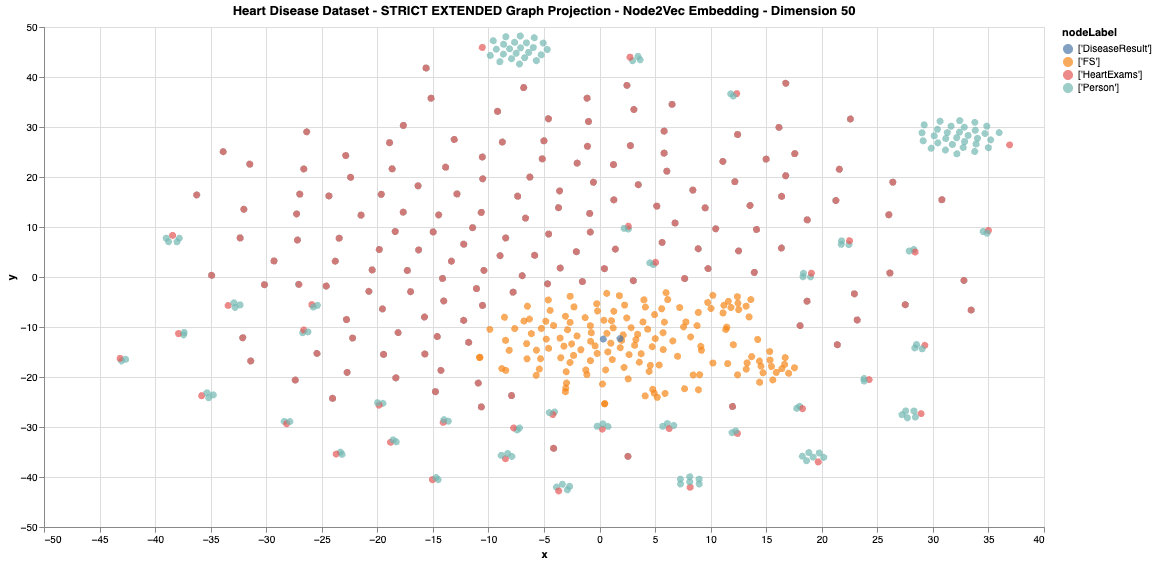}
      \label{fig:extended_node2vec_50}}
  \subfigure[]{
      \includegraphics[width=0.45\linewidth]{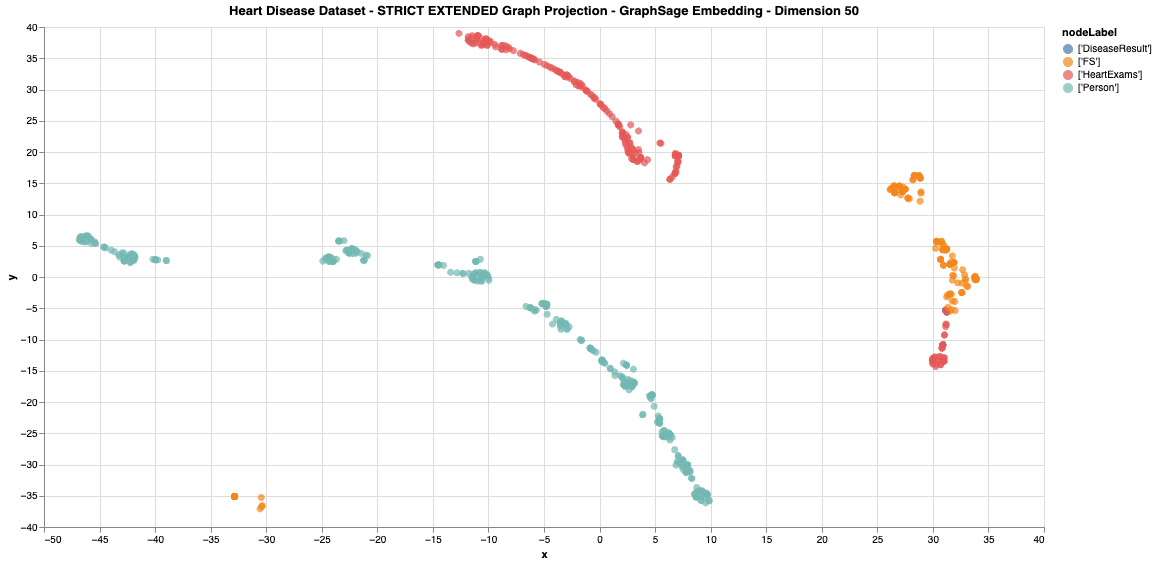}
      \label{fig:extended_graph_sage_50}}
    \subfigure[]{
      \includegraphics[width=0.45\linewidth]{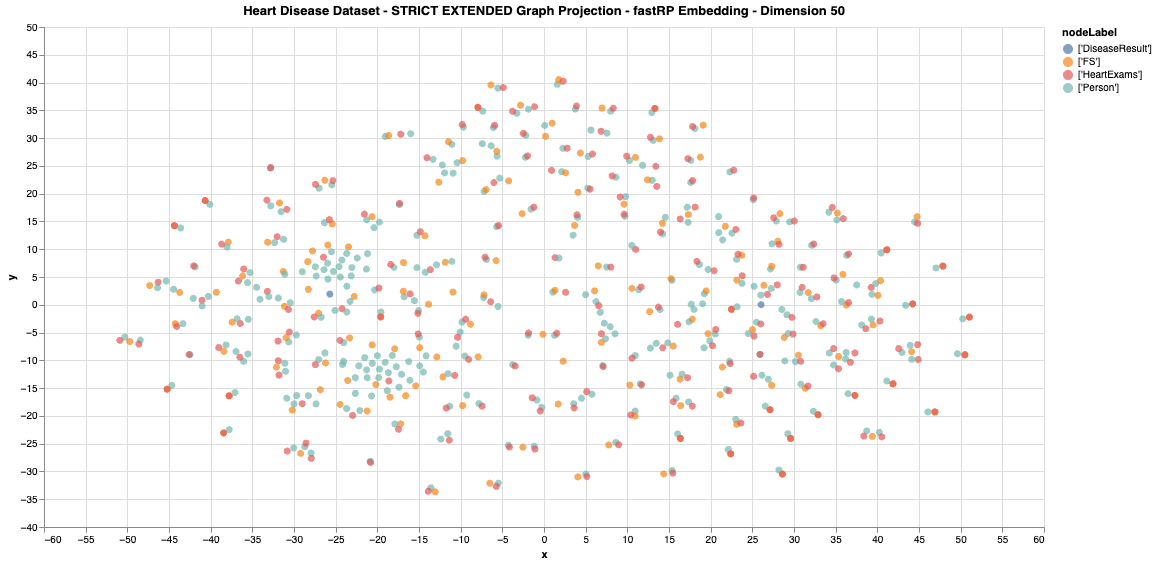}
      \label{fig:extended_fastrp_50}}    
  
  \caption{Strict Extended Graph Projection - Node2Vec, GraphSAGE and FastRP - Dimension 50}\label{fig:50dim}
\end{figure*}

Figures \ref{fig:extended_node2vec_50}, \ref{fig:extended_graph_sage_50}, and \ref{fig:extended_fastrp_50} 
illustrate the embeddings generated for the Strict-Extended Graph Projection using the Node2Vec, GraphSAGE, and 
FastRP algorithms with 50 dimensions for all nodes. Due to the focus on a subgraph, this representation 
exhibits a reduced diversity of node types. As depicted in Figure \ref{fig:10dim}, the GraphSAGE embeddings 
demonstrate superior discriminative results, leading to more accurate node classification compared to 
Node2Vec and FastRP embeddings. While FastRP embeddings exhibit slightly better discriminative capabilities 
than Node2Vec embeddings, it's evident that a dimension of 50 does not significantly alter the quality of the 
embeddings. Instead, the type of projection plays a insteresting role in determining the 
representativeness of the embeddings for classification purposes. In this particular case, GraphSAGE 
embeddings prove to be more suitable for effective classification.

\begin{figure*}[hpbt]
  \centering
   \subfigure[]{%
      \includegraphics[width=0.45\linewidth]{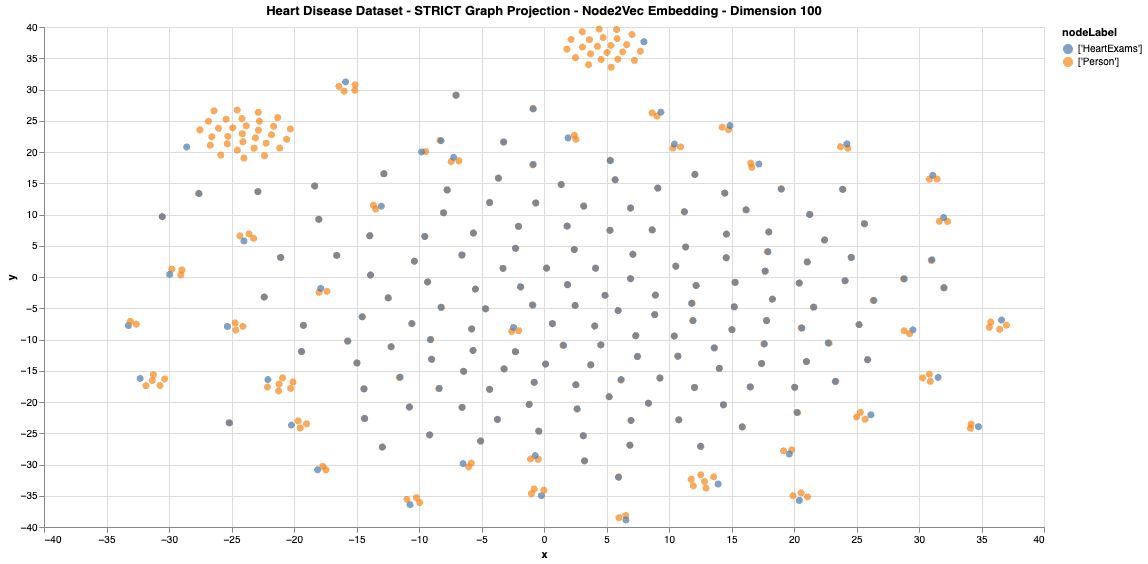}
      \label{fig:strict_node2vec_100}}
  \subfigure[]{
      \includegraphics[width=0.45\linewidth]{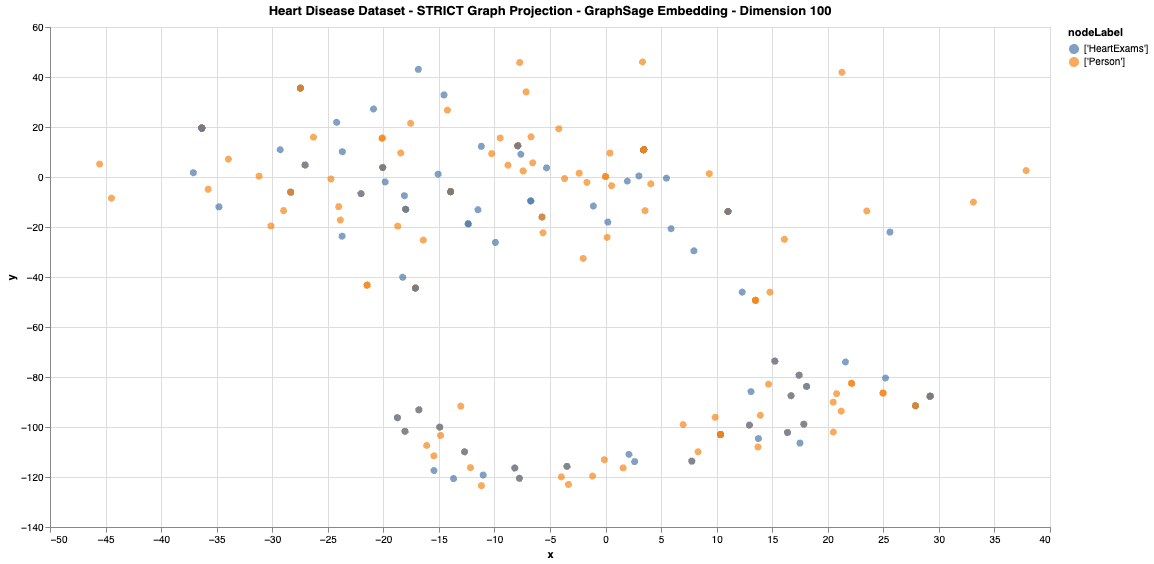}
      \label{fig:strict_graph_sage_100}}
    \subfigure[]{
      \includegraphics[width=0.45\linewidth]{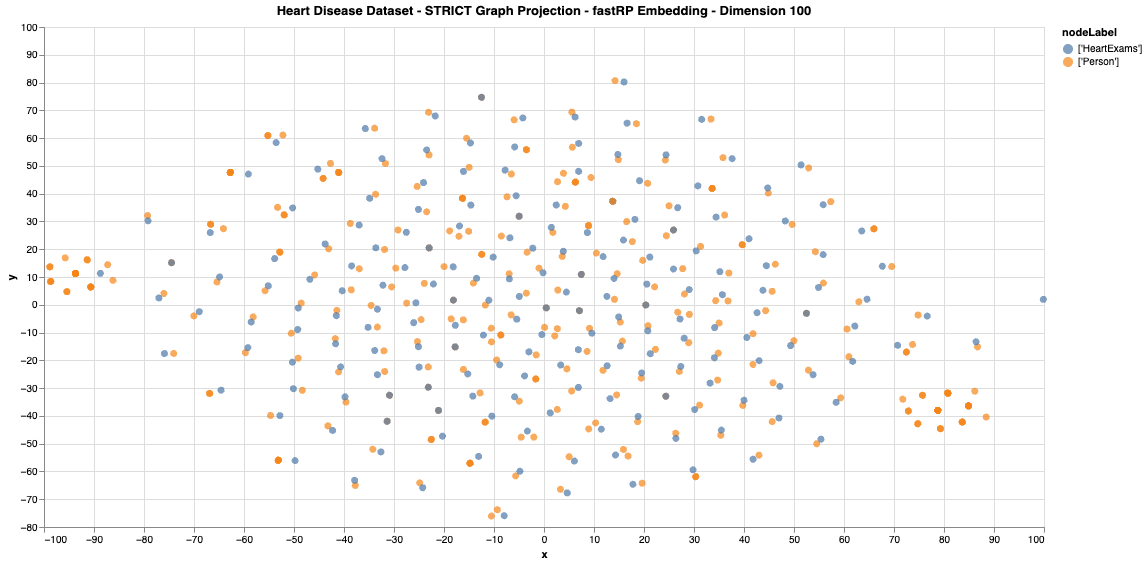}
      \label{fig:strict_fastrp_100}}    
  
  \caption{Strict Graph Projection - Node2Vec, GraphSAGE and FastRP - Dimension 100}\label{fig:100dim}
\end{figure*}

Figures \ref{fig:strict_node2vec_100}, \ref{fig:strict_graph_sage_100} and \ref{fig:strict_fastrp_100} show the 
embeddings generated for the Strict Graph Projection using the  Node2Vec, GraphSAGE, and FastRP algorithms with
100 dimensions for only to types of nodes. In this representation there are less types off nodes, because only a subgraph 
is considered. Different as seen in Figures  \ref{fig:10dim} and \ref{fig:50dim},  GraphSAGE algorithm does not have 
a better performance compared to the other algorithms. The Node2Vec embeddings are more discriminative than the other embeddings, which can 
group better the 2 types of nodes.

\begin{figure*}[hpbt]
  \centering
   \subfigure[]{%
      \includegraphics[width=0.45\linewidth]{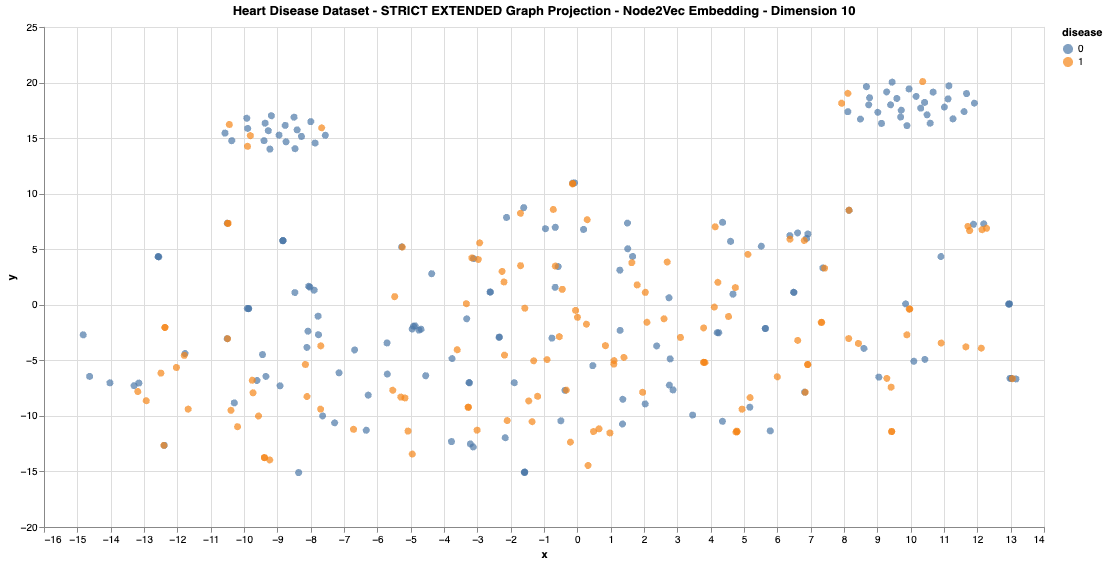}
      \label{fig:extended_node2vec_10_person}}
  \subfigure[]{
      \includegraphics[width=0.45\linewidth]{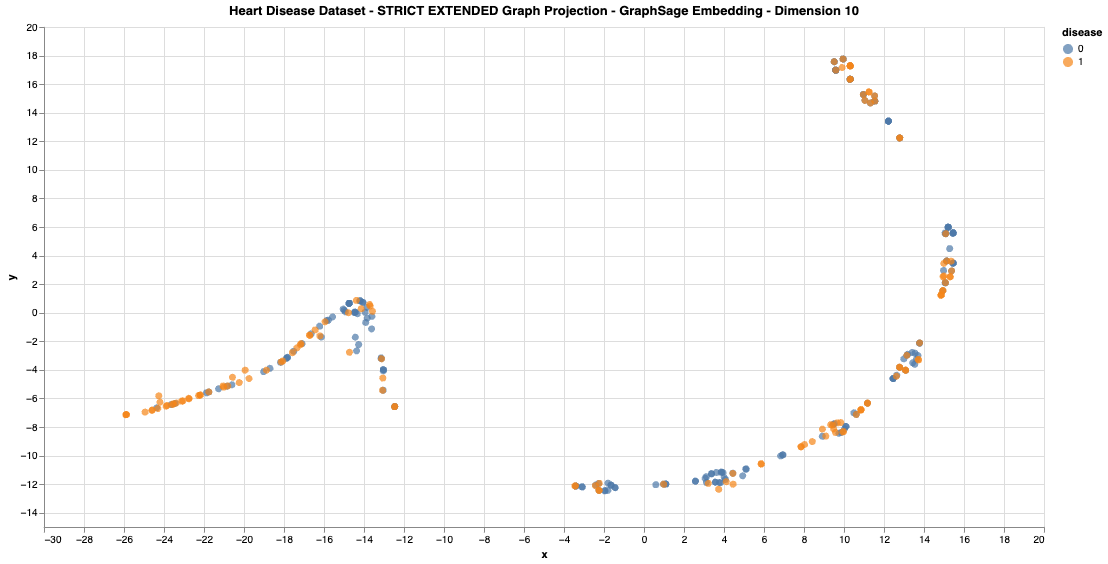}
      \label{fig:extended_graph_sage_10_person}}
    \subfigure[]{
      \includegraphics[width=0.45\linewidth]{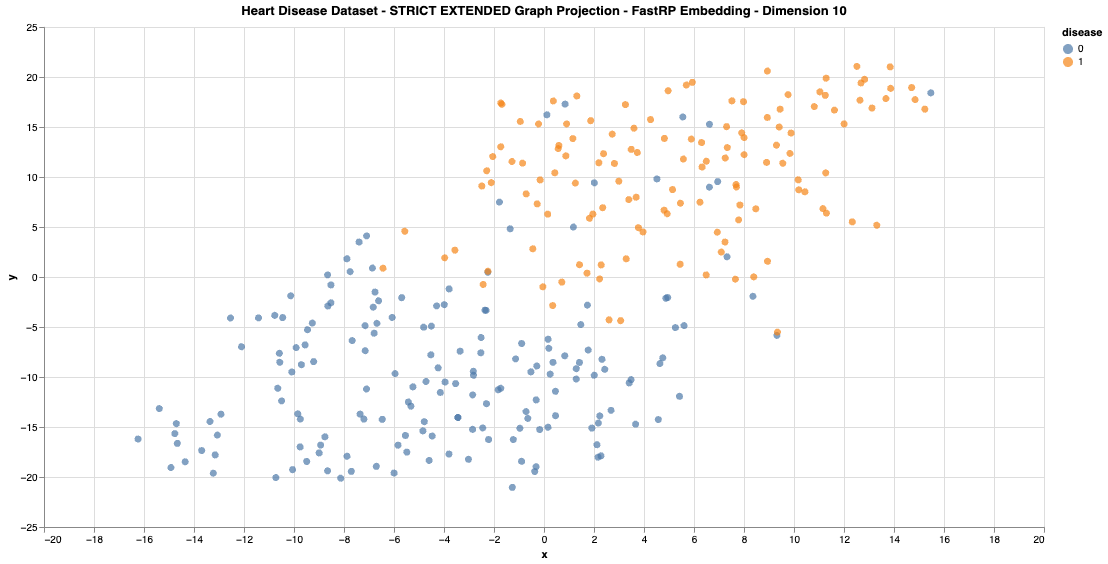}
      \label{fig:extended_fastrp_10_person}}    
  
  \caption{Strict Extended (Person Node) - Node2Vec, GraphSAGE and FastRP - Dimension 10}\label{fig:10dimperson}
\end{figure*}

Following the analysis of embeddings generated for the Full Graph Projection, Strict Graph Projection, and 
Strict-Extended Graph Projection using Node2Vec, GraphSAGE, and FastRP algorithms with varying dimensions, 
it would be intriguing to investigate whether these embeddings could effectively differentiate between healthy 
and sick patients. After executing all projections and generating embeddings, it's evident that the embeddings 
produced for the Strict-Extended Graph Projection offer particularly promising results.

Figure \ref{fig:10dimperson} presents the embeddings generated for the Strict-Extended Graph Projection using the Node2Vec, 
GraphSAGE, and FastRP algorithms with 10 dimensions specifically for Person nodes. In this representation, only one node 
type (Person) is present. As illustrated in Figure \ref{fig:10dimperson}, in this particular case, FastRP embeddings 
are more representative for classifying individuals as healthy or sick.

\begin{figure*}[hpbt]
  \centering
   \subfigure[]{%
      \includegraphics[width=0.45\linewidth]{images/emb/disease/extended_fastrp_10_person.png}
      \label{fig:extended_node2vec_10_person}}
  \subfigure[]{
      \includegraphics[width=0.45\linewidth]{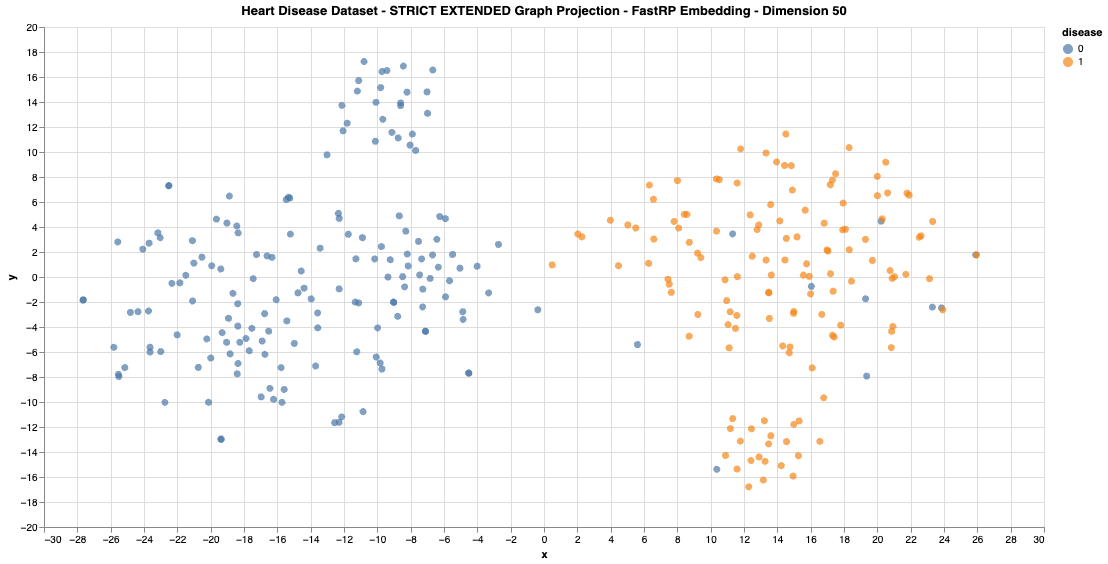}
      \label{fig:extended_fastrp_50_person}}
    \subfigure[]{
      \includegraphics[width=0.45\linewidth]{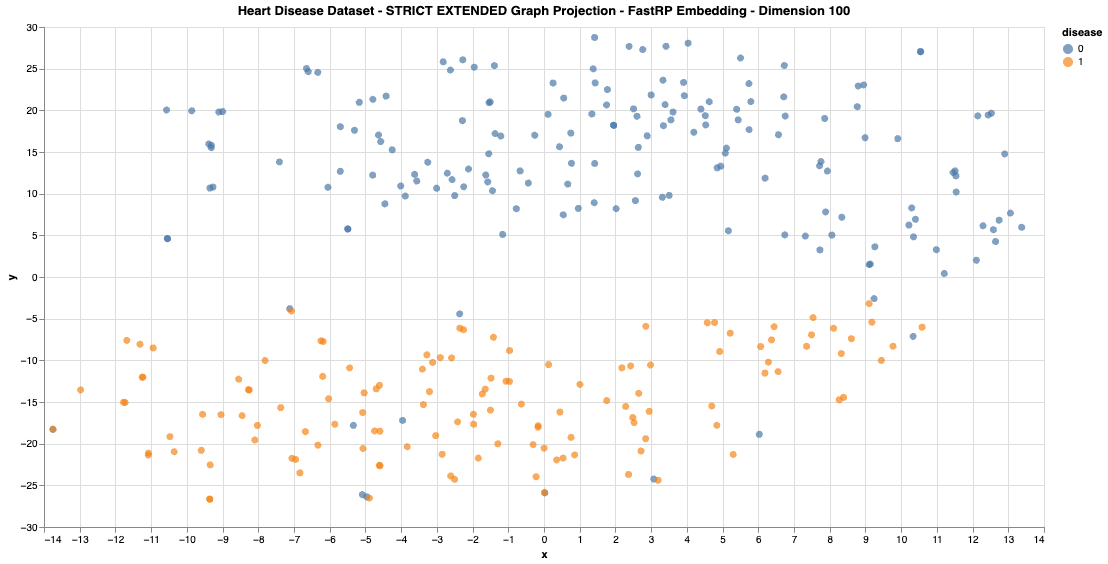}
      \label{fig:extended_fastrp_100_person}}    
  
  \caption{Strict Extended (Person Node) - FastRP - Dimension 10, 50 and 100}\label{fig:10dimpersonfastrp}
\end{figure*}

Following the analysis presented in Figure \ref{fig:10dimperson}, we focused specifically on 
the FastRP embeddings generated for the Strict-Extended Graph Projection, further exploring 
dimensions of 50 and 100 (Subfigures \ref{fig:extended_fastrp_50_person} and \ref{fig:extended_fastrp_100_person}) 
exclusively for Person nodes. This investigation aimed to assess whether the dimensionality could influence the results, 
potentially leading to improved classification or grouping of individuals as healthy or sick.
As it is possible to see in Figure \ref{fig:10dimpersonfastrp} the embeddings generated for the FastRP algorithm 
with 50 and 100 dimensions are more discriminative than the embeddings with 10 dimensions.

The pipeline execution for the Heart Disease dataset demonstrated that the FastRP embeddings generated with 50 and 100 
dimensions yielded superior results for inferring the health status of individuals, distinguishing between healthy and 
sick patients.

\section{Conclusions}

In this work, we proposed a pipeline for predicting information over data graphs. The pipeline consists of several steps,
including data definition and cleaning, query prediction, graph model definition and import, graph projection, embedding generation,  
data visualization, and prediction results. The pipeline was executed for the Heart Disease dataset with different configurations of projections and embeddings.  
The results show that the embeddings generated for the FastRP algorithm with 50 and 100 dimensions are more discriminative for classifying people that are sick.  
The pipeline can be used to analyze the impact of different projections and embeddings on the prediction results for other datasets.  
The pipeline can also be used to analyze the impact of different machine learning models on the prediction results for graph data.  
This work is only a first step in the development of a predictive query-based pipeline for graph data, and there are many opportunities for future work.  

Some future works we can consider are:
\begin{itemize}
  \item Analyze the impact of different machine learning models on the prediction results for graph data;
  \item Analyze the impact of different types of projections and embeddings on:
  \begin{itemize}
    \item the prediction results for other datasets;
    \item the prediction results for other types of machine learning models;
    \item the prediction results for other types of queries;
    \item the prediction results for other types of graph data.
  \end{itemize}
\end{itemize}  

\bibliographystyle{alpha}
\bibliography{biblio}

\end{document}